\documentclass[
 reprint,
 superscriptaddress,
 nofootinbib,
 notitlepage,
 amsmath,amssymb,
 aps,
 prd,
 floatfix,
]{revtex4-2}

\pdfoutput=1
\usepackage{graphicx}
\usepackage{dcolumn}
\usepackage{bm}
\usepackage{aas_macros}
\usepackage{color}

\newcommand{\mr}{\mathrm}
\newcommand{\bmt}{\bm{\theta}}
\newcommand{\bma}{\bm{\alpha}}
\newcommand{\bml}{\bm{\ell}}
\newcommand{\hT}{\hat{T}}

\begin{document}

\title{Cosmic microwave background lensing with optimal convergence and shear estimators}

\author{Hong-Ming Zhu}
\affiliation{%
 Canadian Institute for Theoretical Astrophysics, University of Toronto, 60 St. George Street, Toronto, Ontario M5S 3H8, Canada
}%
\affiliation{%
 Berkeley Center for Cosmological Physics and Department of Physics, University of California, Berkeley, California 94720, USA
}%
\affiliation{
 Lawrence Berkeley National Laboratory, 1 Cyclotron Road, Berkeley, California 94720, USA
}%
 
\author{Ue-Li Pen}%
\affiliation{%
 Canadian Institute for Theoretical Astrophysics, University of Toronto, 60 St. George Street, Toronto, Ontario M5S 3H8, Canada
}%
\affiliation{%
 Dunlap Institute for Astronomy and Astrophysics, University of Toronto, 50 St. George Street, Toronto, Ontario M5S 3H4, Canada
}%
\affiliation{%
 Canadian Institute for Advanced Research, CIFAR Program in Gravitation and Cosmology, \\
 Toronto, Ontario M5G 1M1, Canada
}%
\affiliation{%
 Perimeter Institute for Theoretical Physics, 31 Caroline Street North, Waterloo, Ontario N2L 2Y5, Canada
}%

\date{\today}

\begin{abstract}
We present the optimal convergence and shear estimators for lensing
reconstruction from the cosmic microwave background temperature field.
This generalizes the deflection estimator, is sensitive to non-lensing
modes, provides internal consistency checks, and is always at least as
optimal.  Previously, these estimators were only known in the squeezed
limit. 
This paper  decomposes convergence and shear fields into cosine and
sine waves and the lensed correlation function is then Taylor expanded
in the wave amplitudes. 
Maximizing the likelihood function gives the optimal estimators for
the convergence and shear fields. 

This method has the potential to improve the lensing reconstruction of
the cosmic microwave background polarization field: the shear and
convergence can be optimally combined to form a deflection estimator,
or used separately to separate non-lensing modes, or utilize lensing of
non-Gaussian secondary foregrounds.
\end{abstract}

\maketitle

\section{\label{sec:introduction}Introduction}

Gravitational lensing of the cosmic microwave background (CMB) has
been recognized as a powerful probe of the large-scale structure of
the Universe \cite{2006PhR...429....1L,2016arXiv161002743A}.  Weak
lensing has the advantage of directly tracing the matter distribution
in the Universe and thus avoids the uncertainties with the relation
between the galaxy and matter distributions.  Precision measurements
of the CMB lensing can be used to constrain the neutrino masses, dark
energy, primordial non-Gaussianity, and the halo masses, etc
\cite{2016arXiv161002743A}.  CMB lensing has been measured at high
significance by current surveys (see e.g., Planck
\cite{2020A&A...641A...8P}, ACT \cite{2020arXiv200401139D}, SPTpol
\cite{2019ApJ...884...70W}, and others) and future ground CMB
experiments will continue to improve the measurements substantially
(e.g., Simons Observatory \cite{2019JCAP...02..056A}, CMB-S4
\cite{2016arXiv161002743A}).

The performance of CMB lensing reconstruction depends on the
algorithms used to extract the lensing signal from the observed CMB
map.  The optimal quadratic deflection estimator is constructed by
expanding the observed CMB temperature and polarization to linear
order in the lensing deflection angle
\cite{2001ApJ...557L..79H,2002ApJ...574..566H}.  Within this linear
approximation, the quadratic estimator gives an optimal estimate for
the lensing deflection field.

The deflection field reconstructs a displacement vector, i.e. two
numbers at each point.  In lensing, three numbers are observable: one
convergence and two shears.  In single plane lensing, all effects are
a single scalar degree of freedom.  In post-Born lensing, a curl, or
B-mode can be generated.  In contrast, gravitational waves are
distinguishable from both scalar and B
lensing\cite{2003PhRvL..91b1301D}, which illustrates information lost
in the deflection estimation procedure.  This paper will recover this
lost information.


The quadratic estimator is constructed based on the linear order
lensing effect on the CMB and thus can be biased and suboptimal due to
the higher order terms
\cite{2003PhRvD..67d3001H,2003PhRvD..68h3002H,2003PhRvD..67l3507K}.
Therefore, the maximum likelihood estimator has been first proposed in
Refs.~\cite{2003PhRvD..67d3001H,2003PhRvD..68h3002H} and further
explored in Refs.~\cite{2017PhRvD..96f3510C,2019PhRvD.100b3509M}.
However, the maximum likelihood estimators are generally very
difficult to compute and have to be evaluated iteratively.  In
addition, the estimators based on the deflection angle are susceptible
to the foreground contamination and can have a significant lensing
bias for reconstruction with the CMB temperature
\cite{2014ApJ...786...13V,2018PhRvD..97b3512F,2018PhRvD..98b3534M,2019PhRvL.122r1301S}.

The convergence and shear estimators has been proposed by considering
the distortion of local CMB features from the large-scale lensing
modes in Refs.~\cite{1999PhRvL..82.2636S,1999PhRvD..59l3507Z}, and the
optimal weights in the long wavelength (squeezed) limit have been
derived in
Refs.~\cite{2008MNRAS.388.1819L,2012PhRvD..85d3016B,2018JCAP...01..034P}.
With the independent convergence and shear and information, it is
possible to separate the lensing signal from the lensing bias due to
the extragalactic foregrounds \cite{2019PhRvL.122r1301S}.  Unlike the
optimal quadratic deflection estimator, the local convergence and
shear estimators are only optimal on large scales and become
non-optimal on smaller scales.

The multipole estimators can in principle reach the optimality of the quadratic deflection estimators \cite{2019PhRvL.122r1301S}, but it is also very difficult to apply to the real CMB data.

In this paper, we present the optimal convergence and shear estimators for the lensing reconstruction from CMB temperature map.
We expand the lensed CMB correlation function to linear order in the convergence and shear fields in position space.
The optimal estimators are given by the solution to the maximum likelihood function.
The minimum variance combination of the convergence and shear estimators is equally optimal as the quadratic estimator on small scales and even better than the quadratic estimator on large scales, which is consistent with the results of the maximum likelihood analysis presented in Ref.~\cite{2003PhRvD..67d3001H}.

This paper is organized as follows.
In Sec.~\ref{sec:formalism}, we introduce the new formalism for describing CMB lensing.
In Sec.~\ref{sec:likelihood_analysis}, we describe the maximum likelihood estimator.
In Sec.~\ref{sec:results}, we test the performance of the estimators in simulations and show the numerical results.
We discuss the future development and conclude in Sec.~\ref{sec:discussion}.

\section{\label{sec:formalism}Formalism}

The convergence and shear describe the differential stretching of
structures in the sky, analogous to the metric in general relativity.
Lensing is a special case of a metric that results from a coordinate
change of Euclidean space.  In this case, one can describe the metric
by a lensing transformation.  An arbitrary metric can contain
intrinsic curvature, making it unreducable to Euclidean space.  This a
full lensing estimator must be constructed in curvilinear space, an
attribute which has slowed down attempts to implement this apart from
the squeezed limit.


In this section, we introduce the nonlocal description for weak lensing where the relative deflection between two points is expressed as an integral over the convergence and shear fields along the unperturbed path.
Then we can Taylor expand the lensed CMB temperature covariance using the convergence and shear to linear order with negligible higher order terms.

\subsection{CMB lensing}

Weak lensing of the CMB photons by the intervening matter distribution remaps the CMB temperature field by the deflection field $\bma=\nabla\phi$ as
\begin{equation}
    \tilde{T}(\bmt)=T(\bmt+\nabla\phi)=T(\bmt)+\nabla\phi\cdot\nabla T(\bmt)+\cdots,
\end{equation}
where $\bmt$ is the direction on the sky, $\tilde{T}$ is the lensed CMB temperature field, $T$ is the unlensed CMB temperature field, and $\phi$ is the lensing potential.

The transformation matrix for lensing remapping from the observed coordinate to the source coordinate, $\bmt^S=\bmt+\nabla\phi$, can be usefully decomposed into convergence and shear as
\begin{equation}
    M^{i}_{\phantom{i}j}=\frac{\partial\theta^i_S}{\partial\theta^j}=\left(\begin{array}{cc}
    1-\kappa-\gamma_1     & -\gamma_2 \\
    -\gamma_2     & 1-\kappa+\gamma_1
    \end{array}\right),
    \label{eq:matrix}
\end{equation}
where $\kappa=-(\phi_{,11}+\phi_{,22})/2$ is the convergence, $\gamma_1=-(\phi_{,11}-\phi_{,22})/2$ and $\gamma_2=-\phi_{,12}$ are the two components of the shear.
Note that here $\phi_{,ij}$ denotes $\partial^2\phi/\partial\theta^i\partial\theta^j$.
For weak gravitational lensing, the convergence and shear fields should be much smaller than unity; therefore, the transformation matrix $M$ is invertible everywhere.

The local transformation defined by $M$ describes the deformation of a feature on last scattering surface of infinitesimal angular size $d\theta$.
The convergence and shear estimators can be constructed from local estimates of the anisotropic  CMB power spectrum at the sky position $\bmt$, but are only optimal in the long wavelength limit.

Therefore, instead of the differential variation of the deflection angle at a position $\bmt$, we want to describe the relative deflection between two points at a finite distance on the sky.
This describes the nonlocal effect of convergence and shear on the distance between two points on the last scattering surface instead of the deformation of a feature in the primary CMB temperature of infinitesimal angular size which is local on the sky.

\subsection{Lensing metric}

The lensing remapping from the observed coordinate to the source coordinate can be described by the matrix $M$ at the sky position.
The invariant distance $ds$ between a point $\bmt_S$ and a neighbouring point $\bmt_S+d\bmt_S$ on the last scattering surface is given by 
\begin{equation}
    ds^2=\delta_{ij}d\theta_S^i d\theta_S^j,
    \label{eq:ds_unlensed}
\end{equation}
which is invariant under the lensing remapping. 
In the observed coordinate, we have 
\begin{equation}
    ds^2=g_{\mu\nu}d\theta^\mu d\theta^\nu,
    \label{eq:ds_lensed}
\end{equation}
 where $g_{\mu\nu}$ is the lensing metric tensor.
Combining Eq.~(\ref{eq:ds_unlensed}) and Eq.~(\ref{eq:ds_lensed}), we have
 \begin{equation}
    \delta_{ij}d\theta_S^id\theta_S^j=g_{\mu\nu}d\theta^\mu d\theta^\nu.
\end{equation}
Using Eq.~(\ref{eq:matrix}), we obtain
\begin{equation}
    \delta_{ij}M^{i}_{\phantom{i}\mu}M^{j}_{\phantom{j}\nu}d\theta^\mu d\theta^\nu=g_{\mu\nu}d\theta^\mu d\theta^\nu,
\end{equation}
for any $d\theta^\mu$ and $d\theta^\nu$, implying
\begin{equation}
    g_{\mu\nu}=\delta_{ij}M^i_{\phantom{i}\mu}M^j_{\phantom{j}\nu}.
\end{equation}
Keeping the linear order terms, we have
\begin{equation}
    g_{\mu\nu}=\left(\begin{array}{cc}
    1-2\kappa-2\gamma_1     & -2\gamma_2 \\
    -2\gamma_2     & 1-2\kappa+2\gamma_1
    \end{array}\right).
    \label{eq:metric}
\end{equation}
In the regime of weak lensing where the lensing distortion is small, it should be a valid approximation to neglect the higher order terms in convergence and shear.

The distance between two points $A$ and $B$ on the last-scattering surface is given by 
\begin{equation}
    S_{AB}=\int_A^B ds=\int_A^B \sqrt{g_{\mu\nu}d\theta^\mu d\theta^\nu}.
    \label{eq:S_AB}
\end{equation}
We wish to write the distance between unlensed positions of $A$ and $B$ in terms of the observed coordinates with the lensing metric.
However, this can not be integrated analytically if we do not know the form of convergence and shear fields in the lensing metric.

Let us first consider the convergence field.
In terms of the Fourier components, we have
\begin{equation}
    \kappa(\bm{\theta})=\int\frac{d^2\ell}{(2\pi)^2}\kappa(\bm{\ell})e^{i\bm{\ell}\cdot\bm{\theta}},
\end{equation}
where $\kappa(-\bm{\ell})=\kappa^*(\bm{\ell})$ since $\kappa(\bm{\theta})$ is a real scalar field.
We only need to consider a half plane in Fourier space, $\ell>0$ and $0\leq\phi_\ell<\pi$.
Thus
\begin{eqnarray}
    \kappa(\bm{\theta})&=&\int_{0\leq\phi_\ell<\pi}\frac{d^2\ell}{(2\pi)^2}\left[\kappa(\bm{\ell})e^{i\bm{\ell}\cdot\bm{\theta}}+\kappa^*(\bm{\ell})e^{-i\bm{\ell}\cdot\bm{\theta}}\right].
\end{eqnarray}
Decompose the complex exponentials into sine and cosine functions, we derive
\begin{equation}
    \kappa(\bm{\theta})=\frac{1}{L^2}\sum_{0\leq\phi_\ell<\pi}\left[2\kappa^r(\bm{\ell})\cos(\bm{\ell}\cdot\bm{\theta})-2\kappa^i(\bm{\ell})\sin(\bm{\ell}\cdot\bm{\theta})\right],
\end{equation}
where $L$ is the size of the periodic box, $\kappa^r(\bm{\ell})$ and $\kappa^i(\bm{\ell})$ are real and imaginary parts of the complex Fourier coefficient $\kappa(\bm{\ell})$.

We write the convergence field as a linear combination of sine and cosine waves, with appropriate weights.
We can easily estimate the magnitude of the coefficient from its variance
\begin{equation}
    \frac{\langle2\kappa^r(\bm{\ell})2\kappa^r(\bm{\ell})\rangle}{L^4}=\frac{2C^{\kappa\kappa}_\ell}{L^2}\;.
\end{equation}
For a $10^\circ$ square patch of sky, $L=\pi/180\approx0.174$.
Given that $C^{\kappa\kappa}$ has a maximum of approximately $2\times10^{-7}$, we find the a standard deviation of $\sim2.57\times10^{-3}$, which is significantly less than one.
The cumulative variance of convergence from all scales is given by $\langle\kappa(\bmt)\kappa(\bmt)\rangle=\int \ell d\ell/(2\pi)C_\ell^{\kappa\kappa}$.
We find that the rms convergence to the last scattering surface is about $6\%$ and $7\%$ for $\ell_{\mr{max}}=3000$ and $5000$; therefore, the higher order term proportional to $\mathcal{O}(\kappa^2)$ is less than one percent.  
From this we verify that the linear order lensing Taylor expansion in convergence and shear is a good approximation \cite{2008MNRAS.388.1819L}.

For the lensing shear fields, we also have
\begin{equation}
    \gamma_1(\bm{\theta})=\frac{1}{L^2}\sum_{0\leq\phi_\ell<\pi}\left[2\gamma^r_{1}(\bm{\ell})\cos(\bm{\ell}\cdot\bm{\theta})-2\gamma^i_{1}(\bm{\ell})\sin(\bm{\ell}\cdot\bm{\theta})\right],
\end{equation}
and
\begin{equation}
    \gamma_2(\bm{\theta})=\frac{1}{L^2}\sum_{0\leq\phi_\ell<\pi}\left[2\gamma^r_2(\bm{\ell})\cos(\bm{\ell}\cdot\bm{\theta})-2\gamma^i_2(\bm{\ell})\sin(\bm{\ell}\cdot\bm{\theta})\right].
\end{equation}
In the following, we will use the dimensionless coefficients $\kappa^c(\bm{\ell})=2\kappa^r(\bm{\ell})/L^2$ and $\kappa^s(\bm{\ell})=2\kappa^i(\bm{\ell})/L^2$, and similarly for the shear fields, where the superscript denote the weights of cosine and sine waves.

\subsubsection{Convergence}

Let us start with just one cosine wave of the convergence field, $\kappa(\bmt)=\kappa^c(\bm{\ell})\cos(\bm{\ell}\cdot\bm{\theta})$.
Now the lensing metric tensor takes the form 
\begin{equation}
    g_{\mu\nu}(\bm{\theta})=\left(\begin{array}{cc}
    1-2\kappa(\bm{\theta}) & 0 \\
    0 & 1-2\kappa(\bm{\theta})
    \end{array}\right),
\end{equation}
and
\begin{equation}
    S_{AB}=\int_A^B\sqrt{1-2\kappa(\bmt)}\sqrt{(d\theta^1)^2+(d\theta^2)^2}\;.
    \label{eq:S_AB_kap}
\end{equation}
Without loss of generality, we can take the wave vector $\bm{\ell}=(\ell_1,0)$ along the $\theta^1$ axis.
Then we have the convergence field $\kappa(\bmt)=\kappa^c(\ell_1,0)\cos(\ell_1\theta^1)$, varying only in the $\theta^1$ direction.
We approximate the first term in the integral to linear order, $\sqrt{1-2\kappa(\bmt)}\approx1-\kappa(\bmt)$ and integrate along the unperturbed path connecting the lensed positions of $A$ and $B$ in the observed coordinates,
\begin{equation}
    S_{AB}=\vartheta\left[1-\kappa^c(\bm{\ell})M_{\kappa^c}^{\bml}(\bm{\theta}_A,\bm{\theta}_B)\right],
\end{equation}
where $\vartheta$ is the magnitude of the relative position vector $\bm{\vartheta}=\bmt_B-\bmt_A$, the relative distance between the lensed positions of points $A$ and $B$, and the matrix
\begin{equation}
    M_{\kappa^c}^{\bml}(\bm{\theta}_A,\bm{\theta}_B)=\frac{\sin(\ell_1\theta^1_B)-\sin(\ell_1\theta^1_A)}{\ell_1\theta_B^1-\ell_1\theta_A^1}.
\end{equation}
To illustrate how convergence changes the relative deflection between two points, we can consider a few extreme examples.
When the amplitude of the cosine wave $\kappa^c(\ell_1,0)$ equals zero, we have $S_{AB}=\vartheta$; the relative distance between the lensed positions of $A$ and $B$ is the same as the unlensed positions.
When $\vartheta^1=2\pi n/\ell_1$ and $n$ is an integer, $M_{\kappa^c}^{\bml}(\bmt_A,\bmt_B)$ is zero; here the convergence effect cancels since it is a sinusoidal function.
When $\vartheta^1$ is much smaller than the wavelength $2\pi/\ell_1$, $M_{\kappa_c}^{\bml}(\bmt_A,\bmt_B)$ approaches $\cos(\ell_1\theta^1_A)$, i.e., the value of convergence at position $A$.
When $\vartheta^1$ is much larger than the wavelength $2\pi/\ell_1$, we have $M_{\kappa^c}^{\bml}(\bmt_A,\bmt_B)\ll1$, where a near cancellation occurs between the two positions.

For a cosine convergence wave $\kappa(\bmt)=\kappa^c(\bml)\cos(\bml\cdot\bmt)$ in any direction, we have
\begin{equation}
    M_{\kappa^c}^{\bml}(\bm{\theta}^A,\bm{\theta}^B)=\frac{\sin(\bml\cdot\bmt_B)-\sin(\bml\cdot\bmt_A)}{\bml\cdot\bmt_B-\bml\cdot\bmt_A},
\end{equation}
where the difference is just replacing $\ell_1\theta^1$ by $\bml\cdot\bmt$.
The inner product $\bml\cdot\bmt$ is the product of the wavenumber $\ell$ and the projection of the vector $\bmt$ in the direction of the wave vector $\bml$.

Similarly, for a sine wave $\kappa(\bmt)=-\kappa^s(\bml)\sin(\bml\cdot\bmt)$, we obtain
\begin{equation}
    S_{AB}=\vartheta\left[1-\kappa^s(\bml)M_{\kappa^s}^{\bml}(\bm{\theta}_A,\bm{\theta}_B)\right],
\end{equation}
where
\begin{equation}
    M_{\kappa^s}^{\bm{\ell}}(\bm{\theta}_A,\bm{\theta}_B)=\frac{\cos(\bml\cdot\bmt_B)-\cos(\bml\cdot\bmt_A)}{\bml\cdot\bmt_B-\bml\cdot\bmt_A}.
\end{equation}

\subsubsection{Shear 1}

For the $\gamma_1$ component of the lensing shear field, the lensing metric can be written as
\begin{equation}
    g_{\mu\nu}(\bm{\theta})=\left(\begin{array}{cc}
    1-2\gamma_1(\bm{\theta}) & 0 \\
    0 & 1+2\gamma_1(\bm{\theta})
    \end{array}\right),
\end{equation}
and the unlensed distance between $A$ and $B$ is given by the observed coordinates,
\begin{equation}
    S_{AB}=\int_A^B\sqrt{(1-2\gamma_1(\bmt))(d\theta^1)^2+(1+2\gamma_1(\bmt))(d\theta^2)^2}\;.
\end{equation}
For a cosine or sine wave of the $\gamma_1$ shear field, $\gamma_1(\bmt)=\gamma_1^c(\bmt)\cos(\bml\cdot\bmt)$ or $\gamma_1(\bmt)=\gamma_1^s(\bmt)\sin(\bml\cdot\bmt)$,
we have
\begin{equation}
    S_{AB}=\vartheta\left[1-\gamma_1^{c/s}(\bml)M_{\gamma_1^{c/s}}^{\bm{\ell}}(\bm{\theta}^A,\bm{\theta}^B)\right],
\end{equation}
where
\begin{equation}
    M_{\gamma_1^c}^{\bml}(\bm{\theta}^A,\bm{\theta}^B)=\frac{(\vartheta^1)^2-(\vartheta^2)^2}{(\vartheta^1)^2+(\vartheta^2)^2}\frac{\sin(\bml\cdot\bmt_B)-\sin(\bml\cdot\bmt_A)}{\bml\cdot\bmt_B-\bml\cdot\bmt_A}.
\end{equation}
and
\begin{equation}
    M_{\gamma_1^s}^{\bm{\ell}}(\bm{\theta}_A,\bm{\theta}_B)=\frac{(\vartheta^1)^2-(\vartheta^2)^2}{(\vartheta^1)^2+(\vartheta^2)^2}\frac{\cos(\bml\cdot\bmt_B)-\cos(\bml\cdot\bmt_A)}{\bml\cdot\bmt_B-\bml\cdot\bmt_A}.
\end{equation}
The factor before the sinusoidal part accounts for the anisotropic nature of the $\gamma_1$ shear field.
When the slope of $\bm{\vartheta}$ is $1$ or $-1$, $\vartheta^1=\vartheta^2$ or $\vartheta^1=-\vartheta^2$, the distance between $A$ and $B$ is invariant under the mapping by $\gamma_1$ shear field.
The $\gamma_1$ shear does not induce variation along these two directions.

\subsubsection{Shear 2}

For the $\gamma_2$ component of the shear field, we have the lensing metric tensor
\begin{equation}
    g_{\mu\nu}(\bm{\theta})=\left(\begin{array}{cc}
    1 & -2\gamma_2(\bm{\theta}) \\
    -2\gamma_2(\bm{\theta}) & 1
    \end{array}\right),
\end{equation}
and the unlensed distance between $A$ and $B$ written in the observed coordinates,
\begin{equation}
    S_{AB}=\int_A^B\sqrt{(d\theta^1)^2+(d\theta^2)^2-2\times 2\gamma_2(\bmt))d\theta^1d\theta^2}\;.
\end{equation}
For a cosine or sine wave of the $\gamma_2$ shear field, $\gamma_2(\bmt)=\gamma_2^c(\bmt)\cos(\bml\cdot\bmt)$ or $\gamma_2(\bmt)=\gamma_2^s(\bmt)\sin(\bml\cdot\bmt)$,
we have
\begin{equation}
    S_{AB}=\vartheta\left[1-\gamma_2^{c/s}(\bml)M_{\gamma_2^{c/s}}^{\bm{\ell}}(\bm{\theta}^A,\bm{\theta}^B)\right],
\end{equation}
where
\begin{equation}
    M_{\gamma_2^c}^{\bml}(\bm{\theta}^A,\bm{\theta}^B)=\frac{2\vartheta^1\vartheta^2}{(\vartheta^1)^2+(\vartheta^2)^2}\frac{\sin(\bml\cdot\bmt_B)-\sin(\bml\cdot\bmt_A)}{\bml\cdot\bmt_B-\bml\cdot\bmt_A}.
\end{equation}
and
\begin{equation}
    M_{\gamma_2^s}^{\bm{\ell}}(\bm{\theta}_A,\bm{\theta}_B)=\frac{2\vartheta^1\vartheta^2}{(\vartheta^1)^2+(\vartheta^2)^2}\frac{\cos(\bml\cdot\bmt_B)-\cos(\bml\cdot\bmt_A)}{\bml\cdot\bmt_B-\bml\cdot\bmt_A}.
\end{equation}
Here, the prefactor of the sinusoidal part reflects the anisotropic nature of the $\gamma_2$ shear field.
When $\bm{\vartheta}$ is along the $\theta^1$ axis or $\theta^2$ axis, the distance between $A$ and $B$ is invariant under the mapping by $\gamma_2$ shear.

\subsection{CMB correlation function}

Lensing remaps the CMB temperature field on the sky and thus changes the correlation function.
The lensed CMB temperature correlation function is given by
\begin{equation}
    C^{\tilde{T}\tilde{T}}(\bm{\theta}_A,\bm{\theta}_B)=C^{TT}(S_{AB}),
\label{eq:corr}
\end{equation}
where $C^{TT}$ is the correlation function of the unlensed CMB temperature, which only depends on the separation between points.
Combining all the convergence and shear waves, the distance between the unlensed positions of $A$ and $B$ is  
\begin{equation}
    S_{AB}=\vartheta\left(1-\sum_\alpha p_\alpha M_{\alpha}(\bmt_A,\bmt_B)\right),
\end{equation}
where $\vartheta$ is the observed distance between two points on the sky, and $p_\alpha$ denotes the lensing parameters, $\kappa^c(\bml)$, $\kappa^s(\bml)$, $\gamma_1^c(\bml)$, $\gamma_1^s(\bml)$, $\gamma_2^c(\bml)$, $\gamma_2^s(\bml)$.
Here we include the $\bml$ dependence in $p_\alpha$ for brevity.
We can approximate the covariance of the lensed CMB temperature to linear order in convergence and shear:
\begin{equation}
    C^{\tilde{T}\tilde{T}}(\bm{\theta}_A,\bm{\theta}_B)=C^{TT}(\vartheta)+\sum_{\alpha}p_\alpha C^{TT}_{\alpha}(\bm{\theta}_A,\bm{\theta}_B),
    \label{eq:cf}
\end{equation}
where the first derivative
\begin{equation}
    C_{\alpha}^{TT}(\bm{\theta}_A,\bm{\theta}_B)=-\frac{\partial C^{TT}(\vartheta)}{\partial\ln\vartheta}M_{\alpha}(\bm{\theta}_A,\bm{\theta}_B).
\end{equation}
Lensing breaks the spherical symmetry and translation invariance and the lensed correlation becomes anisotropic and position dependent on the sky.
The $\mathcal{O}(\kappa^2)$ terms in the lensed covariance is at the percent level. 
Thus the lensed covariance is well approximated by the linear order Taylor expansion in convergence and shear.

\section{\label{sec:likelihood_analysis}Likelihood analysis}

The likelihood function for gravitational lensing retains all the information provided by the observations.
The optimal estimator is given by maximizing the likelihood function to the lensing parameters, the values of the lensing potential or the convergence and shear fields. 
However, the maximum likelihood estimator $\hat{\phi}$ for the lensing potential $\phi$ is very difficult to compute since it is a nonlinear function of the lensing potential and has to be solved iteratively \cite{2003PhRvD..67d3001H}.

Lensing of CMB breaks the statistical isotropy, which is manifested in real space by an orientation and position dependent correlation function $C^{\tilde{T}\tilde{T}}(\bmt_A,\bmt_B)$. 
Maximizing the likelihood to the lensing parameters, in this case the amplitudes of sine and cosine waves ($\kappa^c$, $\kappa^s$, $\gamma_1^c$, etc), we have the optimal estimators for convergence and shear fields.
Within the linear approximation, Eq.~(\ref{eq:cf}), the maximum likelihood estimator reduces to a set of linear algebra operations, for example the matrix multiplication, matrix inverse, matrix trace computation, etc, which are more computational tractable, instead of the maximum likelihood estimator $\hat{\phi}$ where higher order terms in the covariance are important \cite{2003PhRvD..67d3001H}.  

In Sec.~\ref{ssec:likelihood}, we introduce the likelihood function for CMB lensing.
In Sec.~\ref{ssec:likelihood_estimator}, we maximize the likelihood function and derive the estimators for convergence and shear. 
In the derivations below, we will largely follow Ref.~\cite{2003PhRvD..67d3001H}.

\subsection{\label{ssec:likelihood}Likelihood function}

We consider a data set of the measured temperature $\hat{T}(\bmt_i)$ at $N$ positions $\bmt_i$ ($i=1,\dots,N$).
The measured temperature is the sum of the lensed temperature $\tilde{T}$ and the instrument noise $\epsilon$:
\begin{equation}
    \hat{T}(\bmt)=\tilde{T}(\bmt)+\epsilon(\bmt).
\end{equation}
The probability distribution for the measured CMB temperature is describe by a density function $P(\hat{T}\vert\kappa)$, where $\kappa$ is the lensing parameters.
Here, we use the abbreviated notation $\kappa$ for the amplitudes of sine and cosine waves.
The covariance matrix of the measured temperature is
\begin{equation}
    C^{\hat{T}\hat{T}}[\kappa]=C^{\tilde{T}\tilde{T}}[\kappa]+C^{\epsilon\epsilon},
\end{equation}
where $C^{\tilde{T}\tilde{T}}$ is the lensed covariance matrix and $C^{\epsilon\epsilon}$ is the noise covariance matrix.
Here, we assume that both the CMB temperature fluctuations and the instrument noises are Gaussian.
Then the probability density of $\hat{T}$ for a lens configuration $\kappa$ is related to its covariance by a Gaussian function:
\begin{equation}
    P(\hat{T}\vert\kappa)=\frac{1}{(2\pi)^{N/2}\sqrt{\det C^{\hat{T}\hat{T}}}}\exp\left(-\frac{1}{2}\hat{T}^{\dagger}C^{\hat{T}\hat{T}-1}\hat{T}\right).
    \label{eq:probability}
\end{equation}
For simplicity, we will use the negative logarithm $\mathcal{L}$ of the likelihood function in the derivation,
\begin{eqnarray}
    \mathcal{L}[\kappa]&=&-\ln P(\hat{T}\vert\kappa)\nonumber\\
    &=&\frac{1}{2}\hat{T}^\dagger\left(C^{\hat{T}\hat{T}}[\kappa]\right)^{-1}\hat{T}+\frac{1}{2}\ln\det C^{
    \hT\hT}[\kappa].
    \label{eq:likelihood}
\end{eqnarray}

\subsection{\label{ssec:likelihood_estimator}Likelihood-based estimators}

We wish to find a set of lensing parameters $p_\alpha$, which maximize the likelihood function, $\partial\mathcal{L}/\partial p_\alpha=0$.
Differentiating Eq.~(\ref{eq:likelihood}) and using Eq.~(\ref{eq:cf}), we obtain
\begin{equation}
    \hat{p}_\alpha={F}^{-1}_{\alpha\beta}\frac{\hat{T}^{\dagger}C^{\hT\hT-1}C^{TT}_\beta C^{\hT\hT-1}\hat{T}-\mathrm{Tr}[C^{\hT\hT-1}C_\beta^{TT}]}{2},
\end{equation}
where the Fisher matrix
\begin{equation}
    F_{\alpha\beta}=\frac{1}{2}\text{Tr}\left[\left(C^{\hT\hT}\right)^{-1}C^{TT}_{\alpha}\left(C^{\hT\hT}\right)^{-1}C^{TT}_{\beta}\right].
\end{equation}
The maximum likelihood estimator becomes a quadratic estimation process in the linear approximation.
In terms of the maximum likelihood iteration, it corresponds to an initial guess $\kappa=0$, with the covariance evaluated in the case of no lensing.
The estimator converges after a single iteration.
This is only valid when the maximum likelihood point is close to the initial guess $\kappa=0$, which is indeed the case for weak gravitational lensing where the convergence and shear are much smaller than unity.
However, we can still iterate the estimator to get better performance when the nonlinear terms are important.

Remember that the lensing parameters $\kappa^c(\bml)$, $\kappa^s(\bml)$, $\gamma_1^c(\bml)\dots$, are real and imaginary parts of the lensing fields by the factor $2/L^2$.
Then the Fourier modes of the convergence field is simply
\begin{equation}
    \kappa(\bml)=\kappa^r(\bml)+i\kappa^i(\bml)=\kappa^c(\bml)L^2/2+i\kappa^s(\bml)L^2/2,
\end{equation}
and similarly for the shear fields $\gamma_1(\bml)$ and $\gamma_2(\bml)$.

From the two shear components, we can construct the two following linear combinations in Fourier space:
\begin{equation}
    \gamma_E(\bml)=\gamma_1(\bml)\cos2\phi_{\bml}+\gamma_2(\bml)\sin2\phi_{\bml},
\end{equation}
and 
\begin{equation}
    \gamma_B(\bml)=-\gamma_1(\bml)\sin2\phi_{\bml}+\gamma_2(\bml)\cos2\phi_{\bml},
\end{equation}
where $\cos\phi_\ell=\ell^1/\ell$ and $\sin\phi_\ell=\ell^2/\ell$.
The parity-even $\gamma_E$ gives an estimate of the lensing convergence $\kappa$, while the parity-odd $\gamma_B$ estimates the curl part in the lensing remapping, which is usually very small.

The Fisher matrix for the $\hat{\kappa}$ and $\hat{\gamma}_E$ estimator is 
\begin{equation}
    \bm{F}=\left(\begin{array}{cc}
    F_{\kappa\kappa} & F_{\kappa\gamma_E} \\
    F_{\gamma_E\kappa} & F_{\gamma_E\gamma_E}
    \end{array}\right).
\end{equation}
Its inverse gives the variance of the $\kappa$ and $\gamma_E$ estimator,
\begin{equation}
    \bm{N}=\left(\begin{array}{cc}
    N^{\kappa\kappa} & N^{\kappa\gamma_E} \\
    N^{\gamma_E\kappa} & N^{\gamma_E\gamma_E}
    \end{array}\right),
\end{equation}
where $\langle\hat{\kappa}(\bml)\hat{\kappa}(\bml')\rangle=(2\pi^2)\delta^D(\bml+\bml')[C_\ell^{\kappa\kappa}+N^{\kappa\kappa}_\ell]$ and similarly for $\hat{\gamma}_E(\bml)$ and the covariance between them.

\begin{figure*}[htb]
\includegraphics[width=1.0\textwidth]{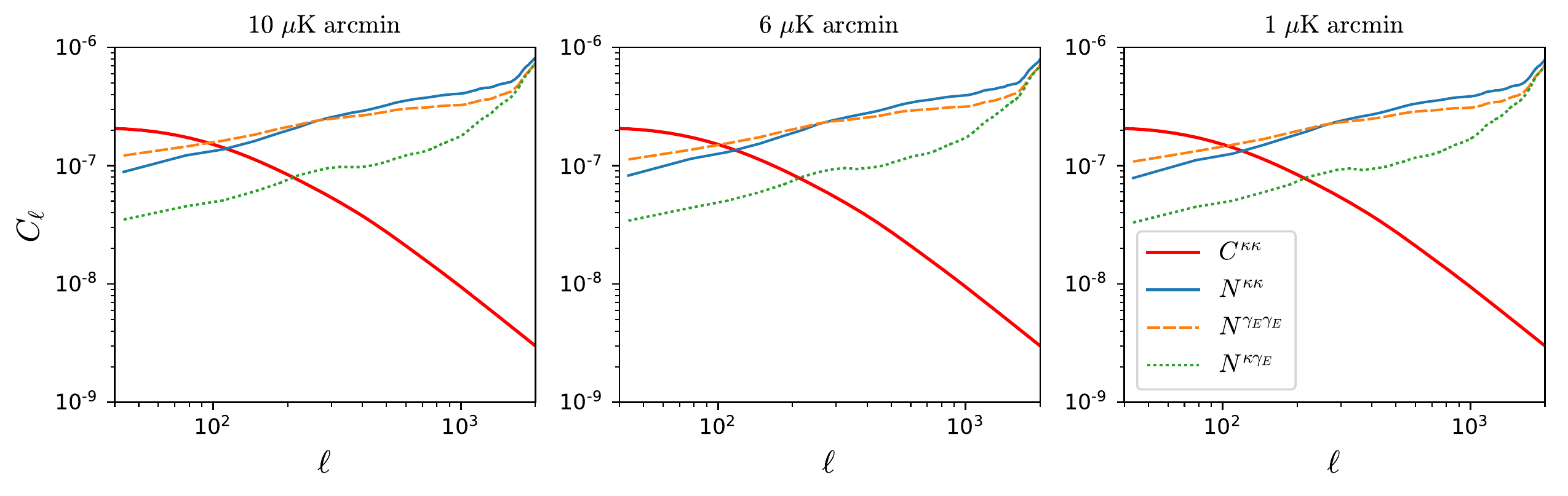}
\vspace{-0.7cm}
\caption{The noise power spectrum for the convergence estimator $N^{\kappa\kappa}$ (solid lines), shear estimator $N^{\gamma_E\gamma_E}$ (dashed lines), and the covariance between these two estimators $N^{\kappa\gamma_E}$ (dotted lines), for experiments with a beam of $1$~arcmin and three noise levels $10~\mu$K~arcmin, $6~\mu$K~arcmin, and $1~\mu$K~arcmin (from left to right).
The thick solid lines show the lensing convergence power spectrum $C^{\kappa\kappa}_\ell$.
Notice that the shear and convergence estimators are nearly independent on large scales and gradually correlated on smaller scales.
\label{fig:f1}}
\end{figure*}

In Fig.~\ref{fig:f1}, 
we plot the noise power spectrum for $\hat{\kappa}$, $\hat{\gamma}_E$ and the covariance between $\hat{\kappa}$ and $\hat{\gamma}_E$, for experiments with a beam of $1'$ and three different noise levels of $10$, $6$, and $1$ $\mu$K~arcmin.
We also plot the lensing convergence power spectrum $C^{\kappa\kappa}_\ell$.
The two estimators are almost independent on large scales and gradually correlated on small scales.

The $\hat{\kappa}$ and $\hat{\gamma}_E$ estimator can be combined to form a minimal variance estimate of the convergence:
\begin{equation}
    \hat{\kappa}_{\mr{MV}}=c_\kappa\hat{\kappa}+c_{\gamma_E}\hat{\gamma}_E,
\end{equation}
where the optimal weights
\begin{equation}
    c_\alpha=\frac{\sum_\beta(\bm{N}^{-1})^{\alpha\beta}}{\sum_{\alpha\beta}(\bm{N}^{-1})^{\alpha\beta}}
\end{equation}
and the noise per mode for the optimal estimator $\hat{\kappa}_{\text{MV}}$ is 
\begin{equation}
    N_{\text{MV}}=\frac{1}{\sum_{\alpha\beta}(\bm{N}^{-1})^{\alpha\beta}}.
\end{equation}

Within the Gaussian approximation and away from the survey boundary or inhomogeneous noise regions, the convergence power spectrum is just the ensemble average of the Fourier coefficients of the convergence field subtracting the noise power spectrum or the inverse of the Fisher matrix \cite{2003PhRvD..67d3001H}.

\section{\label{sec:results}Implementation and results}

To test the performance of the new estimators, we apply the convergence and shear estimators to CMB lensing simulations.
In this section, we consider the power spectrum of the error in the convergence reconstruction and present the convergence power spectrum estimated from the simulated CMB maps.

\subsection{Numerical simulations}

We generate the CMB lensing simulations on a $10^\circ\times10^\circ$ path with pixels of $0.5$~arcmin square on the sky.
The primary CMB temperature field is lensed using the method described in Ref.~\cite{2013MNRAS.435.2040L}.
We consider experiments with three noise levels, $10$, $6$, and $1\ \mu$K~arcmin, and a beam of $1$~arcmin, which roughly corresponds to the noise levels of the ACT experiment \cite{2020arXiv200401139D}, Simons Observatory \cite{2019JCAP...02..056A} and CMB-S4 \cite{2016arXiv161002743A}.
In the analysis, we use the multipole range from $\ell_{\text{min}}=36$ to $\ell_\text{max}=2500$.
The results are averaged over ten independent simulations and the error bars are the $1\sigma$ uncertainty of the scatter between the simulations.

\subsection{Results}

\begin{figure*}[htb]
\includegraphics[width=1.0\textwidth]{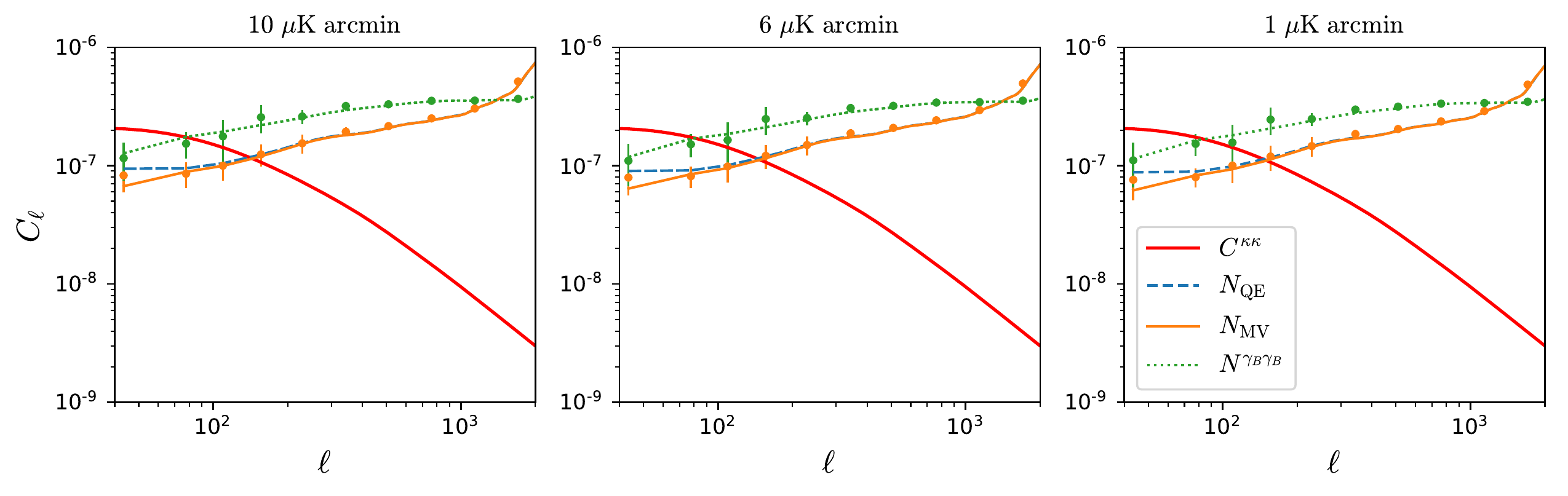}
\vspace{-0.7cm}
\caption{The noise curves for the quadratic deflection estimator (dashed lines) and the minimal variance estimator (solid lines).
The convergence power spectrum is shown in thick solid lines.
The noise power for the $\hat{\gamma}_B$ estimator is also plotted in dotted lines.
The data points show the results from simulations.
The measured power spectra are averaged over ten simulations and the error bars show the rms error between the simulations.
The minimal variance estimator is slightly better on large scales and equally optimal as the quadratic deflection estimator on smaller scales. 
\label{fig:f2}}
\end{figure*}

Figure~\ref{fig:f2} 
shows the noise power spectrum of the minimal variance combination of the convergence and shear estimators for three different white noise levels.
We also plot the noise power spectrum of the quadratic deflection estimator for comparison.
The convergence map errors are measured by computing the difference between input and reconstructed convergence maps, $\hat{\kappa}_{\mr{MV}}-\kappa$.
The power spectra of the error in the convergence reconstruction are shown in Fig.~\ref{fig:f2}.
We find that the numerical results agree well with the theoretical noise curves computed using the Fisher matrix.
The theoretical noise power spectrum of the $\hat{\gamma}_B$ estimator is also plotted and the numerical results are consistent with the theoretical predictions as well.

Therefore, the minimal variance combination of the convergence and shear estimators is equally optimal as the quadratic deflection estimators on smaller scales and is even better on larger scales, which is consistent with the results of the maximum likelihood analysis of the lensing deflection field (see Fig.~4 of Ref.~\cite{2003PhRvD..67d3001H}).
This confirms the validity of the linear approximation made in the lensed covariance Eq.~(\ref{eq:cf}).
However, keep in mind that the convergence and shear estimators can still be iterated to obtain better performance, although for weak lensing we find that the linearized version of the maximum likelihood estimator suffices.

\begin{figure*}[htb]
\includegraphics[width=1.0\textwidth]{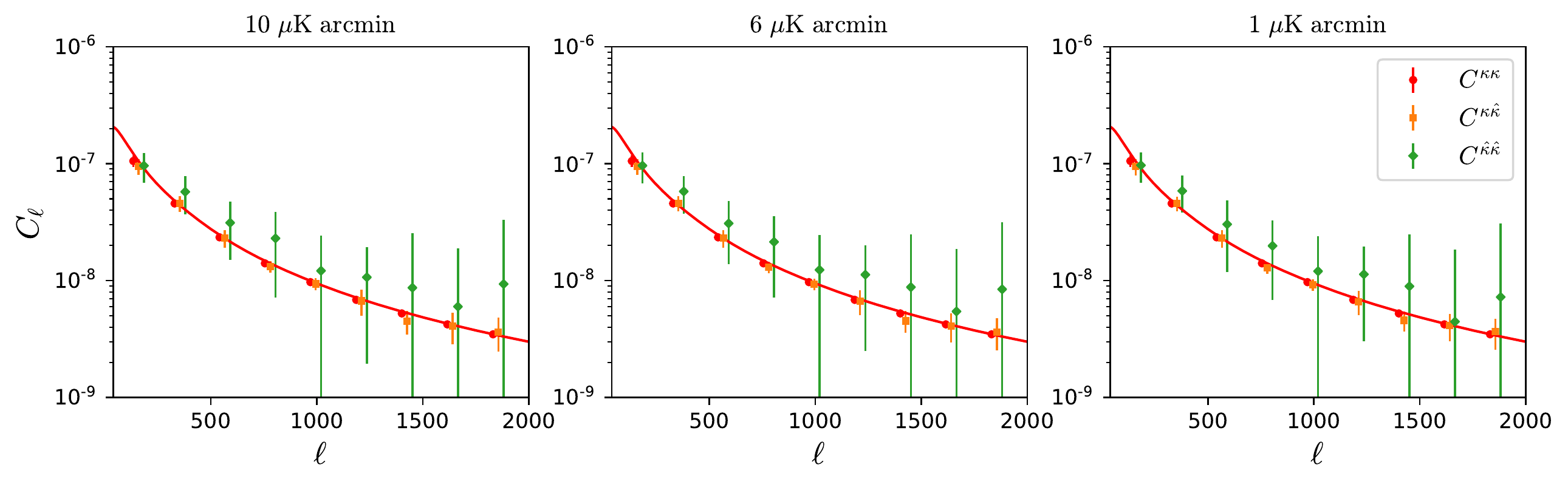}
\vspace{-0.7cm}
\caption{\label{fig:ps}
The power spectrum of the input and reconstructed convergence fields and the cross power spectrum between them.
The solid curves show the theoretical convergence power spectrum.
The data points in each $\ell$  bin are displaced for clarity. 
The measured power spectra are averaged over the ten simulations and the error bars show the scatter between the ten simulations.
The cross power spectra $C^{\kappa\hat{\kappa}}$ agree very well with the input power $C^{\kappa\kappa}$.
The reconstructed convergence power spectrum $C^{\hat{\kappa}\hat{\kappa}}$ agree well with the input power spectrum $C^{\kappa\kappa}$ on large scales and are still consistent with $C^{\kappa\kappa}$ within the $1\sigma$ uncertainty on all scales.
This is mostly due to the very large noise power in lensing reconstruction on smaller scales, where the noise power is tens of times higher than the convergence power spectrum.
}
\end{figure*}

In Fig.~\ref{fig:ps}, we plot the power spectrum of the input and reconstructed convergence fields and the cross power spectrum between them.
The data points are displaced slightly to avoid overlapping.
We find that the cross power spectrum agrees very well with the input power spectrum on all scales, even small scales where the reconstruction noises dominate.
The reconstructed convergence power spectrum agrees well with the input power on large scales and still consistent with the input power within the $1\sigma$ uncertainty.
This is due to the large reconstruction noise on small scales, where the noise is higher than the signal by orders of magnitude.

\section{\label{sec:discussion}Discussion}

In this paper we present the optimal convergence and shear estimators, which improves upon the previous convergence and shear estimators which are only optimal in the long wavelength limit \cite{2008MNRAS.388.1819L,2012PhRvD..85d3016B,2018JCAP...01..034P}.
This is achieved by decomposing the convergence and shear fields using the sine and cosine waves and then expanding the lensed correlation function to linear order in convergence and shear.
Maximizing the likelihood function gives the estimator for the lensing fields.

The methodology to calculate the lensed correlation function is similar to the method to compute the lensed CMB power spectrum, where the lensed correlation is computed in configuration space, ensemble averaged over both the primary CMB temperature fluctuations and the lens configurations, and then computing the inversion to the power spectrum  \cite{1996ApJ...463....1S,2005PhRvD..71j3010C}.
After averaged over an ensemble of lens configurations, the lensed correlation function only depends on the separation between two points; the linear order terms $\sim\mathcal{O}(\kappa)$ vanishes as the expectation value of $\kappa$ is zero and the $\mathcal{O}(\kappa^2)$ terms contribute at the leading order.
However, we are expanding the lensed correlation function for a given lens realization and the lensed correlation function is both anisotropic and position dependent, which is a manifestation of the linear order effect of lensing.
A related discussion is presented in Ref.~\cite{2020A&A...642A.122B}, but is still based on the deflection field instead of the convergence and shear fields considered here.
From the analysis in this paper, we find that the lensing reconstruction from the CMB temperature field is limited by the maximum multipole $\ell_\text{max}$ included in the analysis instead of the experimental noise level.
Increasing the maximum multipole $\ell_\text{max}$ in the analysis will induce lensing biases due to the extragalactic foregrounds \cite{2014ApJ...786...13V,2018PhRvD..97b3512F,2018PhRvD..98b3534M,2019PhRvL.122r1301S}.
The shear estimator is less susceptible to the foregrounds and thus can use a higher $\ell_\text{max}$ \cite{2019PhRvL.122r1301S}.
It is also possible to separate the lensing signal from the foreground biases with the independent shear and convergence information which we plan to explore in the future.

The maximum likelihood method is the same as the previous maximum likelihood analysis for the lensing potential \cite{2003PhRvD..67d3001H,2003PhRvD..68h3002H}.
The difference is that we write the lensed covariance using the convergence and shear fields instead of the deflection angle or the lensing potential as in the previous studies.
Since the linear approximation neglects the second-order and higher order terms in the covariance, the new estimators can still be biased and non-optimal due to terms beyond the linear order.
There is still possibilities to obtain more information from higher order correlation functions.
The new estimators here can be write in the iterative form to perform a nonlinear analysis.
The bias and optimality of the new estimators need to be tested with simulations to assess the validity of the linear approximation as the tests for quadratic deflection estimator (see e.g., Refs.~\cite{2018PhRvD..98l3510B,2019JCAP...10..057F,2018PhRvD..98d3512B} for recent discussions).
We defer a more careful analysis to a future work.
The estimators are constructed in position space instead of Fourier space and can be directly generalized to the analysis with boundaries and inhomogeneous noises. 
 
For non-Gaussian lensing sources, e.g. 21cm\cite{2008MNRAS.388.1819L}
or CIB, or non-Gaussian noise, the convergence and shear estimators
will exhibit non-Gaussian variances.  The separate construction allows
an optimal combination, which improves optimality relative to
deflection estimator, and provides an intrinsic consistency check.

Equipped with a generalized lensing estimator, one might ask how to 
generate a non-trivial non-displacement field.  The simplest case is a
non-Gaussian foreground, e.g. KSZ, CIB, for which this decomposition
provides a broader noise matrix.  In the Gaussian case, say with a
primordial gravitational wave, one would need to specify the
metric. The geodesic distance given by Eq.~(\ref{eq:S_AB}) describes the
distance between pairs of points, and one would need to diagonalize
Eq.~(\ref{eq:corr}) to generate random numbers.  This diagonalization
would no longer be an FFT or spherical harmonic transform.
Nevertheless, it implements a (non-stationary) Gaussian Random field.

\begin{acknowledgments}
This paper is in memory of Suannai, who accompanied the author H.-M. Z. during the preparation of the manuscript.
We thank Simone Ferraro, Emmanuel Schaan, and Martin White for useful discussions and Alexander van Engelen for initial collaborations and sharing the lensing simulations.
We receive support from Natural Sciences and Engineering Research Council of Canada (NSERC) [funding reference number RGPIN-2019-067, 523638-201, CITA 490888-16], Canadian Institute for Advanced Research (CIFAR), Canadian Foundation for Innovation (CFI), Simons Foundation, and Alexander von Humboldt Foundation. 
The simulations are performed on the Sunnyvale computing cluster at CITA and Niagara supercomputer at the SciNet HPC 
Consortium. 
SciNet is funded by the Canada Foundation for Innovation under 
the auspices of Compute Canada; the Government of Ontario; Ontario Research
Fund--Research Excellence; and the University of Toronto.
The Dunlap Institute is funded through an endowment established by the David 
Dunlap family and the University of Toronto.
Research at the Perimeter Institute is supported by the Government of Canada 
through Industry Canada and by the Province of Ontario through the Ministry of 
Research and Innovation.
\end{acknowledgments}

\bibliographystyle{apsrev}
\bibliography{ms}

\end{document}